# IT Complexity Revolution: Intelligent Tools for the Globalised World Development

Andrei Kirilyuk[*] and Mihaela Ulieru

Adaptive Risk Management Laboratory, Faculty of Computer Science
University of New Brunswick, P.O. Box 4400 Fredericton
New Brunswick, E3B 5A3 Canada
`Andrei.Kirilyuk@Gmail.com, ulieru@unb.ca`

**Abstract.** Globalised-civilisation interaction intensity grows exponentially, involving all dimensions and regions of planetary environment. The resulting dynamics of critically high, exploding complexity urgently needs consistent understanding and efficient management. The new, provably universal concept of unreduced dynamic complexity of real interaction processes described here provides the former and can be used as a basis for the latter, in the form of "complexity revolution" in information systems controlling such "critically globalised" civilisation dynamics. We outline the relevant dynamic complexity properties and the ensuing principles of anticipated complexity transition in information and communication systems. We then emphasize key applications of unreduced complexity concept and complexity-driven IT to various aspects of post-industrial civilisation dynamics, including intelligent communication, context-aware information and control systems, reliable genetics, integral medicine, emergent engineering, efficient risk management at the new level of socio-economic development and resulting realistic sustainability.

## 1 Introduction

Exponentially growing power of various interaction processes within increasingly globalised civilisation has surpassed today the characteristic *generalised globalisation* threshold, after which everything becomes related to everything else and the number of unconditionally negligible interaction links tends to zero. A major characteristic of such "truly globalised", strong-interaction system structure and dynamics is that they cannot be a-priori defined - but rather emerge from the "bottom-up" interactions between component systems (and people). This clashes with the traditional systems engineering that approaches the design of hard- and software "with the end in mind", namely by a-priory defining the system and its performance requirements following a "top-down", linear thinking. Given the exponentially huge, always growing number

---

[*] On leave from Institute of Metal Physics, 36 Vernadsky Avenue, 03142 Kiev, Ukraine

of possibilities, such approach cannot be efficient above globalisation threshold, for any linear, basically sequential computing power. It is interesting to note that there is a similar contradiction in fundamental science between predetermined configuration of its "exact-solution", or "perturbative", constructions and much more variable, emergent diversity of real-system structure it is supposed to describe.

In order to cope with this fundamental difficulty and realise the new, *emergent engineering* approach [1], one should first understand the *unreduced interaction dynamics* within a real system, including both controlled natural/human systems and controlling information systems supposed to correctly reproduce main features of controlled system behaviour. Special attention should be given here to processes of new, a-priori unknown structure formation, so that one could eventually replace explicit system design with definition of final purposes and general criteria (that should also be rigorously and universally specified).

The unreduced, real-system interaction description with the necessary properties has been proposed recently for arbitrary kind of system, leading to the *universal concept of dynamic complexity* that was then confirmed by application to various particular systems, from fundamental physics to autonomic communication and information systems [2-6]. We shall outline below major features of those results having strategic importance for IT Revolutions programme (Sect. 2) and showing that to a large degree it can be properly specified in terms of *IT complexity revolution* implying indeed a qualitatively big transition from zero-complexity to high-complexity, intelligent and autonomic ICT structures (Sect. 3). While this *complexity transition* in artificial structures remains a real *challenge* we face today, the properties we should obtain and their origin in natural structures are clearly specified indicating ways to problem solution and providing various intuitive expectations about them with a rigorous basis and quantitative criteria. We conclude by demonstrating, in Sect. 4, inevitability of ICT complexity revolution for key applications, such as reliable genetics, integral medicine and new, complex-dynamic eNetwork of creatively monitored, decentralized and sustainable production processes [7,8].

## 2   Unreduced Interaction Dynamics: Causal Randomness, Emerging Structure and Unified Complexity Definition

Arbitrary interaction configuration, including that of information-exchanging entities within an autonomic eNetworked system, can be *universally* represented by a Hamiltonian-form equation we call here *existence equation* because it simply fixes the fact of many-body interaction within a system, without any special assumption [2-6]:

$$\left\{\sum_{k=0}^{N}\left[h_k(q_k) + \sum_{l>k}^{N} V_{kl}(q_k, q_l)\right]\right\} \Psi(Q) = E \Psi(Q), \tag{1}$$

where $h_k(q_k)$ is the "generalised Hamiltonian" for the *k*-th system component, $q_k$ are its degrees of freedom, $V_{kl}(q_k, q_l)$ is the (arbitrary) interaction potential between the *k*-th and *l*-th components, $\Psi(Q)$ is the system state-function, $Q \equiv \{q_0, q_1, \ldots, q_N\}$, $E$ is the generalised Hamiltonian eigenvalue, and summations include all (*N*) system components. Note that explicitly time-dependent system configurations (with time-dependent interaction potentials) are described by the same equation, up to formal notation change, but here we want to concentrate on time-independent initial configuration, in order to emphasize *emergent* time (and structure) effects.

A usual, perturbative approach to system behaviour analysis involves essential reduction of generally unsolvable (nonintegrable) Eq. (1) to a formally solvable one, but missing many real-system interaction links, such as

$$[h_0(\xi) + \tilde{V}_n(\xi)]\psi_n(\xi) = \eta_n \psi_n(\xi), \tag{2}$$

where $\xi = q_0$ is one of system degrees of freedom, $\psi_n(\xi)$ is a state-function component, and the "mean-field" potential

$$|V_{nn}(\xi)| \lesssim |\tilde{V}_n(\xi)| \lesssim |\sum_{n'} V_{nn'}(\xi)|. \tag{3}$$

It is by choosing a particular configuration of this extremely simplified mean-field potential that one artificially imposes the "expected", predetermined interaction result, any genuine emergence (structure-formation, or self-organisation) effects being practically cut off, irrespective of further "stability analysis" or "peculiar" trajectory behaviour.

However, there is a way to efficiently analyse such "intractable" (practically all real!) interaction problems without their unjustified, creation-killing simplification. It can be done in the (properly extended) framework of so-called optical, or effective, potential method [2-6], where one obtains an equation formally similar to approximate Eq. (2) but being, contrary to the latter, equivalent to the unreduced problem formulation, Eq. (1), in its full complexity:

$$h_0(\xi)\psi_0(\xi) + V_{\text{eff}}(\xi;\eta)\psi_0(\xi) = \eta\psi_0(\xi), \tag{4}$$

where $\eta$ is the problem eigenvalue to be found, while the *effective potential (EP)* $V_{\text{eff}}(\xi;\eta)$ *depends* now, in a very complicated and *highly nonlinear* way, on this eigenvalue and eigenfunction $\psi_0(\xi)$ to be found (details can be found in Refs. [2-6] and other papers cited therein). This EP structure actually contains, now in a more convenient (though always formally "nonintegrable"!) form, the full complexity of all system interactions, their emerging links, possible "round-about ways", etc. It is not difficult to show that this nonlinear, "additional" EP dependence on the eigen-solutions to be found leads to real existence of *multiple*, locally *complete* and therefore *mutually incompatible* problem solutions there where one would expect only one its solution within any perturbative or exact-solution approach of Eqs. (2)-(3). Therefore these mutually incompatible but *equally* real solutions called system *realisations* (they describe its various possible, now *explicitly emerging* configurations) are forced, by the driving interaction itself, to *permanently replace one another*, in a *dynamically (causally) random* order thus defined.

The phenomenon of *dynamic multivaluedness*, or *redundance*, thus *rigorously derived* by the *unreduced* interaction analysis changes dramatically the richness and quality of system behaviour (it should be distinguished from usual "multistability" or "strange attractors" obtained within the reduced, dynamically single-valued analysis equivalent to a mean-field approximation of Eqs. (2)-(3)). First of all, it provides the *universal and strictly intrinsic (dynamic) origin (and the very meaning) of randomness* in *any* real interaction process, which does *not* depend on (formally postulated) time or any other external factors (like "initial conditions" for "diverging trajectories"). It is confirmed by the related *dynamic* definition of *a-priori* event probability, which is not related to any abstract, postulated "event space" but directly follows from the above random change of *dynamically(objectively) equal* realisations:

$$\alpha_r = \frac{N_r}{N_\Re} \ \left( N_r = 1,\ldots,N_\Re; \sum_r N_r = N_\Re \right), \ \sum_r \alpha_r = 1 \ , \tag{5}$$

where $\alpha_r$ is the probability of *dynamic emergence* of *r*-th actually observed realisation containing $N_r$ elementary realisations ($N_r = 1$ for each of these), while $N_\Re$ is the (total) *number of system realisations*. This emergent randomness and probability phenomenon shows, in particular, that any hope for at least theoretically possible regularity of "pure" (properly isolated) and "thoroughly controlled" interaction processes (also in ICT systems) is vain: the origin of randomness is *within* even the formally totally regular, "deterministic" interaction itself (even for zero-uncertainly "initial conditions"!), while any additional "control" attempt is but a new interaction configuration subject to the same, intrinsic randomness.

Directly related to this result is the *universal, dynamic event definition*, which is nothing but each subsequent *realisation (or their dense group) emergence/change* in the process of their permanent "competition" (due to dynamic multivaluedness) within initially "homogeneous" and *time-independent* interaction process. This is the *rigorously specified* process of structure formation, or emergence, or self-organisation, that *cannot* be specified in principle within any usual, *dynamically single-valued* analysis (because the single system realisation just remains identical to itself, without any intrinsic change). It is confirmed by the inevitably related, rigorously derived definition of physically real, emergent, *unceasingly and irreversibly flowing time*. Namely, time acquires an elementary increment as a result of strongly inhomogeneous *realisation change event*, which occurs in our *initially totally timeless* system due to the same dynamic multivaluedness phenomenon. Each real system jump (dynamic reconstruction) from one of its *multiple* realisations (configurations) to another, *randomly chosen* one gives rise to the related *quantum of space* (for a given interaction level), or characteristic (minimum) size, $\Delta x$, directly determined by *eigenvalue spacing* of the unreduced EP problem, Eq. (4), $\Delta x = \Delta_r \eta_i^r$ [2-6], after which the elementary time increment $\Delta t$ is obtained as $\Delta t = \Delta x / v$, where $v$ is the speed of (homogeneous) signal propagation in physically real system "environment" (one of its initial degrees of freedom in Eq. (1)). Unstoppable *time flow* is due to unceasing *realisation change*, while its intrinsic *irreversibility* is due to the *causally random* order of their appearance (time is thus inseparable from dynamic randomness of time-making event emergence). It is evident that this rigorously specified, physically "produced", quantised space and irreversibly flowing time have a *hierarchical structure*, reappearing (emerging) at each new interaction level as relevant-scale entities dynamically "constructed" from effectively homogeneous degrees of freedom of previous, lower level(s).

And finally, one obtains as a unifying result the provably consistent and totally universal definition of *dynamic complexity, C* (and closely related *chaoticity*):

$$C = C(N_\Re), \quad dC/dN_\Re > 0, \quad C(1) = 0, \tag{6}$$

where $N_\Re$ is the system realisation number (determined eventually by the number of its interacting components, $N$ [2-6]) and (dynamic) complexity is universally measured by any *growing* function of realisation number, $C(N_\Re)$, or its rate of change, equal to zero for (actually unrealistic) case of $N_\Re = 1$ (usually $N_\Re \geq N \gg 1$). Note that it is actually the last, totally unrealistic case of zero-complexity, dynamically single-valued interaction result (problem solution) that is invariably considered within usual "models" (including those used in "complexity science", "chaos theory", etc.) that can only provide *effectively zero-dimensional*, point-like "projections" of dynamically multivalued, complex-dynamic behaviour of real systems. Those projections can certainly bear various "signatures" of underlying real-system complexity, but in its strongly and *unpredictably* reduced version. Note, in particular, that $N_\Re$ in Eq. (6) stands for the number of *explicitly obtained* (as a result of unreduced problem solution, Eqs. (1), (4)), mutually incompatible and changing system realisations, rather than the number of arbitrary observed and "countable" entities. Suitable complexity measures include $C(N_\Re) = C_0 \ln(N_\Re)$, $C(N_\Re) = N_\Re - 1$, or $\partial N_\Re / \partial t$. As dynamic multivaluedness underlies both genuine dynamic complexity and causal randomness (see Eq. (5)), we can also define complex behaviour as *chaotic* (dynamically random) one, (dynamical) *chaos* being consistently and universally specified now as (always) causally random process of system *realisation change*, which is the only possible way of any real system or interaction process *existence*. It shows once again that attempts to establish total regularity in any real, complex system cannot be successful in principle, especially in higher-complexity (large and autonomic) systems of our main interest (see also below).

According to intrinsically probabilistic origin of any emerging system configuration, any measured quantity represented by the generalised system density $\rho(Q)$, is obtained as a *causally probabilistic* sum of respective quantity values for individual realisations, $\rho_r(Q)$:

$$\rho(Q) = \sum_{r=1}^{N_\Re} {}^\oplus \rho_r(Q), \tag{7}$$

where detailed expressions for $\rho_r(Q)$ can be obtained within the unreduced EP method using Eq. (4) [2-6] and the sign $\oplus$ designates the special, dynamically probabilistic meaning of the sum. It implies that the observed quantity $\rho(Q)$ permanently, randomly changes, together with system realisation, between its $N_\Re$ possible values $\{\rho_r(Q)\}$ appearing with their *dynamically determined probabilities* $\{\alpha_r\}$ of Eq. (5). The dynamic origin of probability thus obtained means also that, contrary to usual situation, the result of Eqs. (5), (7) remains valid irrespective of the number of observed events (observation time) and, in particular, it is valid for every single realisation (event) emergence (and even *before* any event emergence!).

However, the complete problem solution has even more complicated structure than that of the causally probabilistic sum of Eq. (7) representing only the first level of system dynamics splitting into many incompatible, changing realisations. Each of these realisations generally shows similar internal splitting into second-level, also incompatible and probabilistically changing realisations (under the influence of the same system interaction) and so on, where the total number of such realisation levels can be very large even though not all of them can be practically resolved as such. This general, most complete problem solution (expressed by a *multi-level* probabilistic sum in Eq. (7)) has thus the structure of *dynamically probabilistic fractal* [2-4], whose multi-level hierarchy of ever finer elements is much richer than that of usual fractals because it includes permanent *probabilistic realisation change* at each structure level, providing it with the property of *efficient dynamic adaptability*, or *intelligent (reasonable) behaviour*, observed in *living* organisms.

Useful *power* $P_{\text{real}}$ of that probabilistic fractal dynamics underlying such "magic" properties as high adaptability, autonomy, creativity, intelligence and sustainable development (highly desired for the new ICT and eSocial systems!) is determined by the total number of (fractal) realisations $N_\Re$ (or complexity $C$) that can be estimated as the *number of system link combinations* [3-6]:

$$P_{\text{real}} \propto N_\Re \sim L! \simeq \sqrt{2\pi L}\left(\frac{L}{e}\right)^L \sim L^L \propto C, \qquad (8)$$

where the number of links $L$ is already very large (it can be much greater than the number of system components $N$: for human brain, $N > 10^{10}$, $L > 10^{14} \gg N$). Useful power of corresponding systems with traditionally limited (regular, sequential, linear) operation can at best grow only as $L^\beta \ll P_{\text{real}}$ ($\beta \sim 1$). It is this *exponentially huge* efficiency advantage that explains the above "magic" *qualities* of high-complexity (very large $L$) natural systems (*life, intelligence, consciousness, sustainability*), which can be successfully reproduced and efficiently controlled in man-made, artificial environment only if one "liberates" the involved information (and human!) system dynamics to follow a creative, free-interaction regime.

Every structure-formation, emergent system dynamics resulting from unreduced, complex-dynamic interaction development can also be called *self-organisation*. Although we have shown above that any really emerging, self-organised structure inevitably contains a great deal of dynamic randomness, the latter is usually *confined* to a more or less distinct shape determining the observed system configuration. How can one define then the actual "proportions" of, and the border between, those omnipresent but opposite properties of randomness and order? The detailed analysis of EP method equations (see Eq. (4)) shows [2-6,9] that the onset of strongly irregular regime of *uniform chaos* occurs under the condition of *resonance* between major component processes, such as the internal dynamics of each system component and characteristic interaction transmission between components:

$$\kappa = \frac{\omega_\xi}{\omega_q} \cong 1, \qquad (9)$$

where $\kappa$ is the introduced *chaoticity* parameter, while $\omega_\xi$ and $\omega_q$ are frequencies (or energy-level separations) for the inter-component and intra-component motions, respectively. At $\kappa \ll 1$ (far from resonance) one has a multivalued self-organised or confined-chaos regime (internally chaotic but quasi-regular externally), which becomes the less and less regular as $\kappa$ grows from 0 to 1, until at $\kappa \approx 1$ (resonance condition) the global or uniform chaos sets in, followed by another self-organised regime with an "inverse" system configuration at $\kappa \gg 1$. All multi-level hierarchy of any real system dynamics can be described within this universal classification, where additional complication comes from the fact that there are always many higher-order resonances in the system (describing always present by maybe spatially limited chaoticity). In natural system dynamics the regimes of uniform chaos and self-organisation tend to "reasonably" alternate and coexist, so that the former plays the role of efficient "search of the best development way", while the latter ensures more distinct structure creation as such, both providing detailed realisation of the above high interaction power, Eq. (8), to be reproduced in the next generation of intelligent eNetworks, where the quantitative criterion of Eq. (9) can be useful as a universal guiding line (uniquely related to the above unreduced complexity analysis).

The qualitatively strong, irreversible change and creativity inherent to the unreduced, multivalued system dynamics imply certain *direction* and *purpose* of real interaction processes. These can be specified [2-6,10] as *unceasing* transformation, in *any* interaction process and system dynamics, of a *latent* complexity form of *dynamic information*, $I$ (generalising the notion of "potential energy"), into its *explicit* form of *dynamic entropy*, $S$ (generalising the notion of "kinetic energy"), always occurring so that the *total complexity*, $C$, given by the sum of the two complexity forms, decreasing dynamic information and increasing dynamic entropy, $C = I + S$, remains *unchanged* (it is given by the initial interaction configuration and any its external modification):

$$\Delta C = 0, \quad \Delta S = -\Delta I > 0. \tag{10}$$

Dynamic information describes system potential for new structure/quality creation, while dynamic entropy describes the unreduced dynamic complexity of already created structure. The *universal symmetry of complexity* (law of its conservation and transformation), Eq. (10), provides thus another useful guideline for emergent engineering by rigorously specifying the *universal purpose of natural interaction development*. One can show that all major laws and dynamic equations known from fundamental physics can be reduced to particular cases of this universal symmetry of complexity [2,3,10], which is further evidence in its favour as a guiding rule.

## 3  Complexity Transition: Creative ICT Systems and Emergent Engineering

Rigorous description of unreduced interaction dynamics from the previous section including provably universal concept of complexity can be considered as a necessary basis for the new, *exact* science of intelligent and creative ICT tools and their efficient application to management and development of complex real-world dynamics. That kind of theory provides a *rigorously specified* extension of respective empirical results and intuitive expectations about the next stage of ICT development confirming its now *provably revolutionary* character and specifying its *objectively efficient* content. One should also take into account that such complex information systems are supposed to be used for efficient control of real, high-complexity systems, implying consistent, realistic and universal understanding of controlled system dynamics (otherwise missing) as indispensable condition for their sustainable management.

Major features of unreduced complex dynamics relevant for the IT complexity revolution were outlined in Sect. 2: huge power of unreduced interaction process, its inevitable and purely dynamic randomness, universal classification of major regimes of truly complex dynamics, and the universal symmetry of complexity as the unified law and purpose, as well as guiding line for ICT system design. Now we can further specify these results in terms of several major *principles of complex ICT system operation and design* [6] realising also the ideas of emergent engineering [1].

We start with the *complexity correspondence principle* implying efficient interaction only between systems of *comparable* (unreduced) complexity. Being a direct consequence of complexity conservation law, Eq. (10), it limits the scope of efficient system design to cases that do not contradict that fundamental law and therefore can be realised at least in principle (similar to energy conservation law use in usual, thermo-mechanical machine construction). A major manifestation of the complexity correspondence principle is the "complexity enslavement rule" stating that a higher-complexity system can efficiently control (or enslave) a lower-complexity one, but never the other way around. Therefore, there is no sense to try to obtain efficient (autonomic) control over a complex system using only lower-complexity tools. It can be considered as rigorous substantiation of the importance of the whole IT complexity revolution concept we describe here: in its absence, traditional, zero-complexity systems monitoring complex real-world phenomena can be used only in a strongly non-autonomic way implying essential input from human complexity levels (intelligence), but in today's increasingly "globalised", strong-interaction world efficient application of such man-dominated, "slow" and "subjective" control becomes ever less efficient if not catastrophic.

According to the same principle, interacting similar-complexity systems may easily give rise to strongly chaotic (dynamically multivalued) behaviour that can be both harmful (in situations where one is looking for stability of a basically established configuration) and useful (in situations where essentially new and best possible ways of complexity development should be found). And finally, when one tries to use a very high-complexity system for control over a much lower-complexity one, this control should certainly be possible, but may be practically inefficient for another reason: the high-complexity controller will effectively "replace" (or even suppress) the much lower-complexity but useful process under its "surveillance". This would imply, in particular, that IT complexity revolution should better start and proceed from software/context to hardware/traffic level, rather than in the opposite direction.

These rules of efficient complexity control can be extended to the second major principle of complex ICT management, the *complex-dynamic control principle*. Based on the same universal symmetry of complexity, but now rather its complexity development aspects (see the end of Sect. 2), it states that contrary to ideas of traditional, "fixed" control, the extended, complex-dynamic control implies suitable *complexity development* (i.e. *partially unpredictable change*) as a major condition for efficient system monitoring. It means that, as proven by our unreduced interaction analysis (Sect. 2), controlled dynamics can *not* - and should not - be totally, or even mainly, regular. In other words, suitable degrees of acceptable chaotic change, or even essential structure development, should be provided for truly efficient, failure-proof eControl systems (involving both their ICT and human elements.)

The purpose of such extended, complex-dynamical control becomes thus practically indistinguishable from the general direction of *optimal interaction complexity development* (from dynamic information to dynamic entropy, Sect. 2), as it should be, especially in a "generally globalised" system, taking into account its inseparable interaction structure including both controlling and controlled elements. It shows that properly *creative* monitoring is actually much more *reliable* than traditional, restrictive control, implying once again the necessary essential advance towards complex-dynamic, truly intelligent IT control systems. Combined with the first principle of complexity correspondence and enslavement, this universal guiding line and criterion of complex-dynamic control leads to the ultimate, now uniquely realisable purpose of *sustainable control* providing (unlimited) *global development stability* by way of omnipresent *local creativity*.

Finally, the *unreduced (free) interaction principle* refers to exponentially huge power of unreduced interaction processes, as opposed to much lower, power-law efficiency of traditional, linear (sequential) operation schemes in existing ICT systems (see Eq. (8) in the previous section). The above complexity correspondence principle confirms the evident fact that efficient management of that huge real-interaction power would need equally high, complex-dynamic interaction power of IT systems applied and ever more densely inserted in the tissue of real-world complexity and intelligence. However, such huge power progress needs equally big transition from detailed step-by-step programming in usual ICT approach to monitoring of only general development direction of complexity-entropy growth (by properly specifying complexity-information input). Intrinsic chaoticity should become a normal, useful operation regime. Correspondingly, the huge power of free-interaction dynamics can be realised in the form of dynamically probabilistic fractal (Sect. 2), with specially allocated possibilities of its development. As mentioned above, such natural complexity development tends to occur as irregular, constructive alternation of global chaos and multivalued self-organisation regimes of complex dynamics, the former realising efficient search of optimal development ways and the latter providing a more ordered structure creation as such.

The unified, rigorously substantiated content of these three major principles of complex ICT system operation and design reveals the forthcoming ICT *complexity transition* as the first stage of IT complexity revolution and the beginning of *useful* complexity and chaos role in information-processing systems. This qualitatively big transition will show up as essential power growth and appearance of new features usually attributed to living and intelligent systems. Whereas technical realisation of complexity transition remains an exciting challenge, the objective necessity to meet it follows e.g. from the universal criterion of chaos as being due to system resonances (see Eq. (9)). While a sufficiently low-intensity eNetwork can try to maintain its basic regularity by avoiding major resonances, it becomes impossible for higher network interaction density/intensity due to inevitable overlap of emerging new resonances. Therefore even a limited task of preserving traditional system regularity acquires a nontrivial character and needs application of unreduced complexity description. While everything shows that today we are already quite near that complexity-transition threshold, the unreduced interaction complexity opens also much brighter perspectives of new, generally unlimited network possibilities after successful complexity transition, providing additional motivation to the whole problem study. As strong interaction within the global ICT-human-social system is inevitable in any case and has already been realised in large parts of the world, its often negative aspects of "future shock" [11] (where dense but regular IT systems tend to "impose" their "pathological" linearity to intrinsically complex human thinking) can be replaced by the opposite positive effects only as a result of IT complexity revolution (where human intelligence endows IT systems with a part of its natural complexity).

One relatively easy way to approach the desired complexity transition starting from existing network structure is to attempt a transition from their still hardware- and location-based realisation to intelligent-software- and *knowledge-based* realisation. Such truly knowledge-based networks can be conceived as autonomic systems of interacting and permanently *changing* knowledge (any semantic) structures able to usefully evolve *without* direct intervention of human user (that will instead create and modify general rules and particular preferences of this knowledge interaction process). Efficiency demands for such knowledge-based ICT development will naturally enforce the advent of complex-dynamic operation modes, simply due to complex-dynamic structure of unreduced knowledge content. This major example shows how even a "usual" quality-of-service demand involves complexity transition in ICT system operation and design. This is also the next, equally natural step of emergent engineering [1] realisation in its *autonomic engineering* version, where the omnipresent, real-time (and knowledge-based!) system development constitutes an integral part of its complex, holistic dynamics within any particular application.

## 4  IT Revolutions as Complexity Revolution in Science and Society

The IT complexity revolution substantiated and specified in previous sections as the core of modern IT revolution represents a natural result of ever more interactively and intensely used information and communication systems approaching now the critical point of complexity transition (Sect. 3). This rapidly advancing and still poorly recognised process acquires yet greater importance and support if one takes into account equally big complexity transitions emerging in all key fields of science, technology and global civilisation dynamics constituting together the forthcoming *complexity revolution* in human civilisation development [5]. It becomes evident that all these essentially complex-dynamical applications need complex-dynamic IT tools for their real progress (due to the rigorously substantiated complexity correspondence principle, Sect. 3), while successful development of those tools can only result from complexity science and application progress. It would be not out of place to briefly outline here the key complexity applications asking for IT complexity revolution.

While today's rapid progress of *genetics* and related bio-medical applications is widely acknowledged, emerging serious problems around the unreduced dynamic complexity of bio-chemical systems involved only start appearing in public science discussions (see e.g. [12]). The universal science of complexity clearly specifies the irreducible origin of those problems [3] in terms of exponentially huge power of real interaction processes, with characteristic values of the number of essential interaction links $L \sim 10^{12} - 10^{14}$ in Eq. (8), making any usual (sequential/regular) computing power negligible with respect to practical infinity implied by the estimate of Eq. (8). It follows that only essential progress towards equally rich, truly complex dynamics of information systems used for the causally complete understanding of living organism dynamics can solve this kind of problem and in particular form a solid basis for the truly *reliable genetics*.

Very close to genetics is the extremely popular group of *nano-bio applications*, where the fact of strongly chaotic (multivalued) nano-system dynamics [13] remains practically unknown, despite the evident similarity with the above bio-chemical problems. Here too, the dominating incorrect reduction of nano-system dynamics to regular models can be as harmful as the unreduced complexity analysis can be advantageous, asking for complex-dynamic IT systems as major study tools.

At a higher complexity level of the whole living organism dealt with in *medicine*, it becomes evident that any its correct understanding (absolutely necessary for sustainable progress of extended life quality) should involve suitably complex information system dynamics able to provide a unified complex-dynamical "map" of each individual organism dynamics and development. This is the idea of *integral medicine* [2,3] based on inseparable combination of unreduced complexity science and complex-dynamic IT systems applied. Comparing these natural perspectives and challenges with the reductive and separating approach of usual, mechanistic medicine, it is easy to see the necessity of *bio-medical complexity revolution*.

*Sustainable development* issues represent a further natural extension of the same ideas to ever higher complexity levels of planetary civilisation dynamics involving nevertheless a well-specified conclusion about the necessary unified transition to a superior level of the entire civilisation complexity [5]. This "embracing", *global complexity revolution* can only be avoided in the case of alternative possibility of explicitly degrading, complexity-destroying development branch (that may have already become dangerously real). As knowledge-intense civilisation structure includes, already today, ICT systems as its essential part, the forthcoming global complexity revolution and the next development stage cannot avoid essential use of properly upgraded, complex-dynamical information systems.

One can speak here about *complex eNetworks efficiently controlling decentralized, emergent production and creative consumption systems* [7,8] (in the new sense of "developing" complex-dynamical control of Sect. 3) and effectively replacing today's inefficient financial systems that cannot cope with the real development complexity. Note, by the way, that efficient management of huge volumes of (highly interactive) scientific and technological, all innovation-related information alone would certainly need urgent introduction of knowledge-based, context-aware and thus explicitly complex-dynamic ICT tools.

We shall not discuss here much lower complexity levels of *physical systems*, while noting, however, that even at those *relatively* low complexity levels one is definitely pushed towards the unreduced analysis in terms of real system complexity (instead of conventional zero-complexity "models"), without which "unsolvable" problems and related development impasses accumulate catastrophically in various fields, from particle physics and cosmology to high-temperature superconductivity and nuclear fusion [14], leading to the desperate "end of science".

One should also emphasize a *general-scientific* but practically important meaning of consistent understanding of complex ICT system dynamics. Whereas fields like binary algebra and logics formed a basis for information technology development in a previous epoch, today one deals rather with the problem of (arbitrary) interaction within a real system of many "information bodies" whose causally complete solution just leads to the universal concept of dynamic complexity (Sect. 2). The resulting *complex-dynamic information science* [6] will therefore constitute a rigorous basis for the next stage of fundamental and applied computer science development realising essential progress of scientific knowledge. Complex-dynamic, real-knowledge-based, semantically full and therefore inevitably "uncertain" information entities will replace standard, semantically trivial "bits" from the previous, hardware-oriented level of information science, which is exactly what is needed from the point of view of modern application demands [15-18]. This crucial progress involves a qualitatively deep change in dominating conceptual attitudes and empirical methods as discussed above (Sect. 3).

In conclusion, we have demonstrated that the revolutionary situation in IT development clearly felt today can be consistently and constructively specified as IT complexity revolution implying creation and (practically unlimited) development of qualitatively new ICT systems and applications. Namely, we (1) rigorously proved the inevitability of ICT complexity emergence (Sect. 2), (2) derived major properties of interest of real-system complexity within its universally valid concept (Sect. 2), (3) obtained three major principles of complex-dynamic and intelligent ICT development and related complexity transition (Sect. 3), and (4) demonstrated the application-related need for complex ICT system development, including the global complexity revolution leading to intrinsically sustainable level of social and economic civilisation dynamics (Sect. 4). It is therefore difficult to see another, equally promising and problem-solving alternative to this long-term way of ICT development, with its revolutionary start beginning right now and going to be a major contribution of today's generation to the planetary civilisation development.

## References

1. Doursat, R., Ulieru, M.: Emergent Engineering for the Management of Complex Situations. In: Proceedings of Autonomics 2008, Second ACM International Conference on Autonomic Computing and Communication Systems, Turin, Italy, September 23-25, 2008; http://www.cs.unb.ca/~ulieru/Publications/Ulieru-Autonomics-Final.pdf
2. Kirilyuk, A.P.: Universal Concept of Complexity by the Dynamic Redundancy Paradigm: Causal Randomness, Complete Wave Mechanics, and the Ultimate Unification of Knowledge. Naukova Dumka, Kyiv (1997); http://books.google.com/books?id=V1cmKSRM3EIC. For a non-technical review see also: ArXiv:Physics/9806002
3. Kirilyuk, A.P.: Complex-Dynamical Extension of the Fractal Paradigm and Its Applications in Life Sciences. In: Losa, G.A., Merlini, D., Nonnenmacher, T.F., Weibel, E.R. (eds.)

Fractals in Biology and Medicine, Vol. IV, pp. 233--244. Birkhäuser, Basel (2005); ArXiv:Physics/0502133
4. Kirilyuk, A.P.: Consistent Cosmology, Dynamic Relativity and Causal Quantum Mechanics as Unified Manifestations of the Symmetry of Complexity. Report presented at the Sixth International Conference "Symmetry in Nonlinear Mathematical Physics" (Kiev, 20-26 June 2005); ArXiv:Physics/0601140
5. Kirilyuk, A.P.: Towards Sustainable Future by Transition to the Next Level Civilisation. In: Burdyuzha, V. (ed.) The Future of Life and the Future of Our Civilisation, pp. 411--435. Springer, Dordrecht (2006); ArXiv:Physics/0509234
6. Kirilyuk, A.P.: Unreduced Dynamic Complexity: Towards the Unified Science of Intelligent Communication Networks and Software. In: Gaïti, D. (ed.) Network Control and Engineering for QoS, Security, and Mobility, IV. IFIP, Vol. 229, pp. 1--20. Springer, Boston (2007); ArXiv:Physics/0603132
7. Ulieru, M., Verdon, J.: IT Revolutions in the Industry: From the Command Economy to the eNetworked Industrial Ecosystem. In: Proceedings of the 1st International Workshop on Industrial Ecosystems, IEEE International Conference on Industrial Informatics, Daejoen, Korea, July 13-17 2008;
http://www.cs.unb.ca/~ulieru/Publications/Verdon-paper-formatted.pdf
8. Ulieru, M.: Evolving the 'DNA blueprint' of eNetwork middleware to Control Resilient and Efficient Cyber-Physical Ecosystems. Invited paper at BIONETICS 2007 - 2nd International Conference on Bio-Inspired Models of Network, Information, and Computing Systems, Budapest, Hungary December 10-14, 2007;
http://www.cs.unb.ca/~ulieru/Publications/Ulieru-Bionetics2007.pdf
9. Kirilyuk, A.P.: Dynamically Multivalued Self-Organisation and Probabilistic Structure Formation Processes. Solid State Phenomena 97-98, 21--26 (2004); ArXiv:Physics/0405063
10. Kirilyuk, A.P.: Universal Symmetry of Complexity and Its Manifestations at Different Levels of World Dynamics. Proceedings of Institute of Mathematics of NAS of Ukraine 50, 821--828 (2004); ArXiv:Physics/0404006
11. Toffler, A.: Future shock. Bantam, New York (1984)
12. Gannon, F.: Too complex to comprehend? EMBO Reports 8, 205 (2007)
13. Kirilyuk, A.P.: Complex Dynamics of Real Nanosystems: Fundamental Paradigm for Nanoscience and Nanotechnology. Nanosystems, Nanomaterials, Nanotechnologies 2, 1085--1090 (2004); ArXiv:Physics/0412097
14. Kirilyuk, A.P.: The Last Scientific Revolution. In: López Corredoira, M., Castro Perelman, C. (eds.) Against the Tide: A Critical Review by Scientists of How Physics & Astronomy Get Done, pp. 179--217. Universal Publishers, Boca Raton (2008); ArXiv:0705.4562
15. European Commission R&D Framework Programme, Information Society Technologies, Future and Emerging Technologies, Situated and Autonomic Communications, http://cordis.europa.eu/ist/fet/comms.htm. Links to other related initiatives can be found at http://cordis.europa.eu/ist/fet/areas.htm
16. Bullock, S., Cliff, D.: Complexity and Emergent Behaviour in ICT Systems. Foresight Intelligent Infrastructure Systems Project. UK Office of Science and Technology (2004), http://www.foresight.gov.uk/OurWork/CompletedProjects/IIS/Docs/ComplexityandEmergentBehaviour.asp
17. Di Marzo Serugendo, G., Karageorgos, A., Rana, O.F., Zambonelli, F. (eds.): Engineering Self-Organising Systems: Nature Inspired Approaches to Software Engineering, LNCS Vol. 2977 (Springer Verlag, Berlin, 2004).
18. Smirnov, M. (ed.): Autonomic Communication. First International IFIP Workshop, WAC 2004, Berlin, Germany, October 18-19, 2004, Revised Selected. LNCS, Vol. 3457. Springer, Berlin/Heidelberg (2005)